\documentclass{article}

\usepackage[dvipsnames]{xcolor}
\PassOptionsToPackage{colorlinks,citecolor=RoyalBlue,urlcolor=RoyalBlue,linkcolor=RoyalBlue}{hyperref}
\usepackage{url,fullpage}
\usepackage{pgfplots}
\usepackage{prettyref}
\newrefformat{sec}{Section~\ref{#1}}
\newrefformat{fig}{Figure~\ref{#1}}

\newcommand{\usanon}[2]{#1}

\begin{document}


\title{CS Circles: An In-Browser Python Course for Beginners}


\author{
David Pritchard\footnote{Department of Computer Science, Princeton University, USA, {\tt dp6@cs.princeton.edu}. CS Circles is a project of the Centre for Education in Mathematics and Computing, University of Waterloo, Canada.} \and
Troy Vasiga\footnote{David R.~Cheriton School of Computer Science and the Centre for Education in Mathematics and Computing, University of Waterloo, Canada, {\tt tmjvasiga@cs.uwaterloo.ca}.}}

\maketitle
\begin{abstract}
Computer Science Circles is a free programming website for beginners that is designed to be fun, easy to use, and accessible to the broadest possible audience. We teach Python since it is simple yet powerful, and the course content is well-structured but written in plain language. The website has over one hundred exercises in thirty lesson pages, plus special features to help teachers support their students. It is available in both English and French. We discuss the philosophy behind the course and its design, we describe how it was implemented, and we give statistics on its use.
\end{abstract}
%


\section{Introduction}
The last decade has seen an explosion of web-based software that runs ``in the cloud," jointly with a tremendous improvement in the ability of web browsers to interact with users and servers. Near the end of 2009, we and our colleagues decided to create an interactive introductory programming course using these technologies. The main motivation was that many wonderful online programming judges existed for veteran programmers\footnote{e.g.~TopCoder.com, Sphere Online Judge (\url{spoj.pj}), the ACM judge (\url{acm.uva.es}), USA Computing Olympiad training (\url{train.usaco.org}), the PEG judge (\url{wcipeg.com}), etc.}, but at the same time beginners had little to work with. It is unreasonable to expect all novices to download a compiler, or to re-load a webpage and re-upload a file just to debug their very first program. Computer Science Circles (CS Circles) is our way of filling this gap. It is publicly available at
\begin{center}
\usanon{\url{http://cscircles.cemc.uwaterloo.ca}}{[[Normal URL redacted. You are welcome to visit it at \url{http://bit.ly/K4x83V}; this will reveal our university's name. Beware that the ``Thanks" page there has some of our names listed.]]}
\end{center}
Since opening in the fall of 2011, over 150,000 exercises have been completed on the site, by nearly 7000 registered users. On an average day, the website handles 900 visits and 7000 code submissions. Approximately half of the traffic is from users that are not logged in.\footnote{Summary statistics cover from when CS Circles opened to the time of writing (mid-December 2012). Daily statistics represent averages for the preceding six months.}

We eschew the traditional separation into lessons and sets of exercises. Our position is that such a framework puts students at risk of losing interest mid-lesson, and also splits their attention between exercise comprehension, lesson review, and problem solving when they finally start working hands-on. We think it is pedagogically superior for the exercises to be embedded throughout the lessons: this reduces monotony and keeps the student on a focused progression. Removing page reloads and handling the student/auto-grader communication with asynchronous JavaScript (Ajax) helps keep the experience smooth.

Programming is the ideal topic for computer-based learning: beginners can get immediate detailed feedback on the entirety of their creations from the compiler and auto-grader. But, automatic grading is not a replacement for a real teacher looking over the student's shoulder. To compensate for these issues we have added a human touch by letting students easily send a ``Help" message when they are stuck. Any message sent in this way will also automatically include the code currently in the student's code editor for that problem. We and our colleagues have answered over 1000 such student questions. Additionally, teachers wishing to use CS Circles can easily become the ``guru" of their students, redirecting all questions to the teacher instead. Teachers also can easily track the progress of their students, which mitigates that CS Circles does not check for plagiarism or style.

The rest of the paper is organized as follows. \prettyref{sec:related} discusses related work. \prettyref{sec:philosophy} provides an overview of the CS Circles experience and the philosophy behind it. \prettyref{sec:features} discusses the features of the site, grouped according to the three types of people interacting with the site: students, lesson authors, and teachers. \prettyref{sec:implementation} describes details of the site's implementation. Finally, \prettyref{sec:future} has some suggestions for future work.

\subsection{Related Work}\label{sec:related}
The year 2012 saw a proliferation of MOOCs (massive open online courses) and in particular, the first courses of the companies Coursera (\url{coursera.org}) and Udacity ({\url{udacity.com}). Two main distinctions between these platforms are that Coursera offers each course on a specific schedule, while Udacity allows open enrollment at any time; and that Coursera deals with the full range of university subjects, while Udacity is focused on the sciences, especially computer science. Udacity has an auto-grader, and so do a few Coursera courses. Two other educational websites focus on interactive online programming enrichment in a more independent way, without university partnerships or certification processes. The first, Codecademy (\url{www.codecademy.com}), was launched in October 2011. Initially they only offered JavaScript, but they now offer Python and other languages. The second, Khan Academy (\url{www.khanacademy.org/cs}), has offered online interactive programming lessons in a dialect of JavaScript since August 2012; previously (since 2011) they offered videos on Python. 

The programming environments of both Khan Academy and Codecademy are tailored for learning. Codecademy shows exercises on one side of the browser and a coding window on the other site. Student code is retained in the coding window between steps of a multi-part exercise, which gives the student helpful continuity while progressing through the work. The design of the Khan Academy CS user interface is drastically different from traditional coding environments. Immediately after typing or deleting each character in the program editor, it is re-parsed and re-executed dynamically (with injection to ensure that the current state persists as much as possible). See John Resig's talk~\cite{Resig12} for details, as well as Bret Victor's recent re-popularization of ``responsive programming"~\cite{Victor12}. For example, numeric constants in the code can be changed by dragging a slider, which creates the effect of an animation when there is graphic output. The ubiquituous Khan Academy video takes the form of a narrated replay of the lecturer's code window, which can be interrupted at any time and edited. As an additional feature, syntax/run-time errors come with automatically-generated suggestions on how it can be fixed. These are fantastic tools but beyond the scope of our own project.

Both of these sites have strengths and weaknesses in their pedagogy and interface. For example, Codecademy has some ``hints" containing mandatory directions, and occasionally does not give a reason why a user failed a test; while Khan Academy's ``semicolon monster" occasionally obfuscates an error rather than clarifying it. Nonetheless these sites both represent the state of the art and their UI and curriculum are improving continuously.

In CS Circles we give some of the benefits of both the university-style sites and the more open-ended ones. We have numbered lessons, giving an easy-to-follow linearly structured curriculum, any page of which can be accessed at any time. Since the whole site is entirely focused on Python, every page uses a consistent, well-polished programming interface. One advantage of CS Circles over the others is that it can easily facilitate teachers helping their students when they have specific questions, by using the guru feature. Lesson layout is sequential like the traditional concept of a web ``page." Each lesson, together with its embedded exercises, is laid out in a vertical sequence on a scrollable web page. Despite its old-fashionedness, this has advantages over video/audio tracks: it is simpler for authors and translators, it is easier to review, and it helps students follow at their own speed, including students not fluent in English.

Two other online interactive Python courses date back to 2011. The site \url{pyschools.com} offers 210 exercises arranged in 14 topics, but without overall narrative. The groundbreaking ``How to Think Like a Computer Scientist: Interactive Edition"~\cite{MR11} combines an in-browser Python interpreter (implemented in JavaScript) with Creative Commons-licensed material from Downey's ``Think Python" book~\cite{Downey08}. The user experience with their site is comparable to ours. On the one hand not all of their exercises are auto-gradeable, but on the other hand their auto-grader is able to run without contacting a central server. See also an interactive sequel on algorithms and data structures~\cite{MR12} and a paper on methodology~\cite{MR12conf} by the same authors.

\section{Bird's-eye View}\label{sec:philosophy}
The CS Circles website is ``end-to-end" in the sense that everything is handled through the browser interface: submitting code, seeing the grader's reply, editing (with syntax highlighting), visualization with forward/backward stepping, saving and loading solutions, and asking for help. There is a curriculum to serve as a backbone, but we also make judicious use of flexibility: some lessons are optional, helping each user learn at their own choice of depth, and some lesson groups can be completed in any order, making the course less monotonous.

{\bf Content.} The content begins with a lesson on ``Hello, World!", including a working click-to-run sample program and a crashing one. We continue through variables, function calls, comments, quotes, and all of the fundamentals needed in order to write or understand an elementary program. Most lessons have a narrative where a new feature is motivated and explained, with examples, caveats, and exercises for the user. A few special lessons don't discuss language features but rather practical matters like errors, design, and debugging; and several lessons have only exercises based on material from prior lessons, helping to re-consolidate what the user learned previously. The final lessons discuss basic auto-decryption of Caesar ciphers, recursion, object vs.~value identity, and efficiency. We forced ourselves not to make the lessons too long; 1000 words was a soft upper limit on what was reasonable to convey in a single sitting. The end goal for the course as a whole is to give the user enough knowledge and confidence to understand further information available online. 

The top half of \prettyref{fig:hello} displays the first exercise on the site, which is to write a program that prints out a particular message, in analogy with the {\tt Hello, World!} example. When the user submits their code, it is compiled, executed, and graded server-side. Then, without a page reload, the grading box is updated according to the result of grading. The bottom half of \prettyref{fig:hello} shows the result of a successful submission for this exercise.

\begin{figure}[tb!]
\centering
\includegraphics[width=8cm]{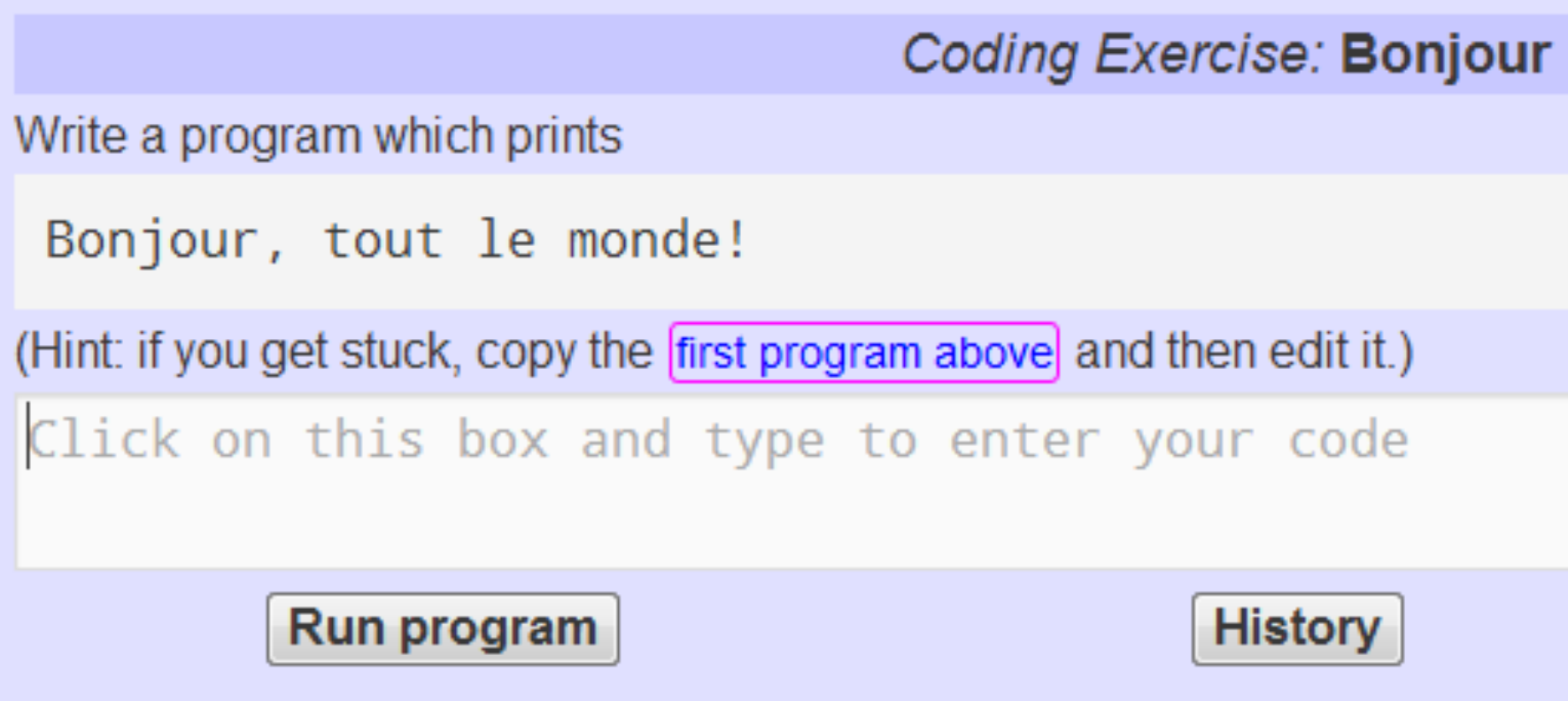} 
\vspace{0em}$$\Downarrow$$

\includegraphics[width=8cm]{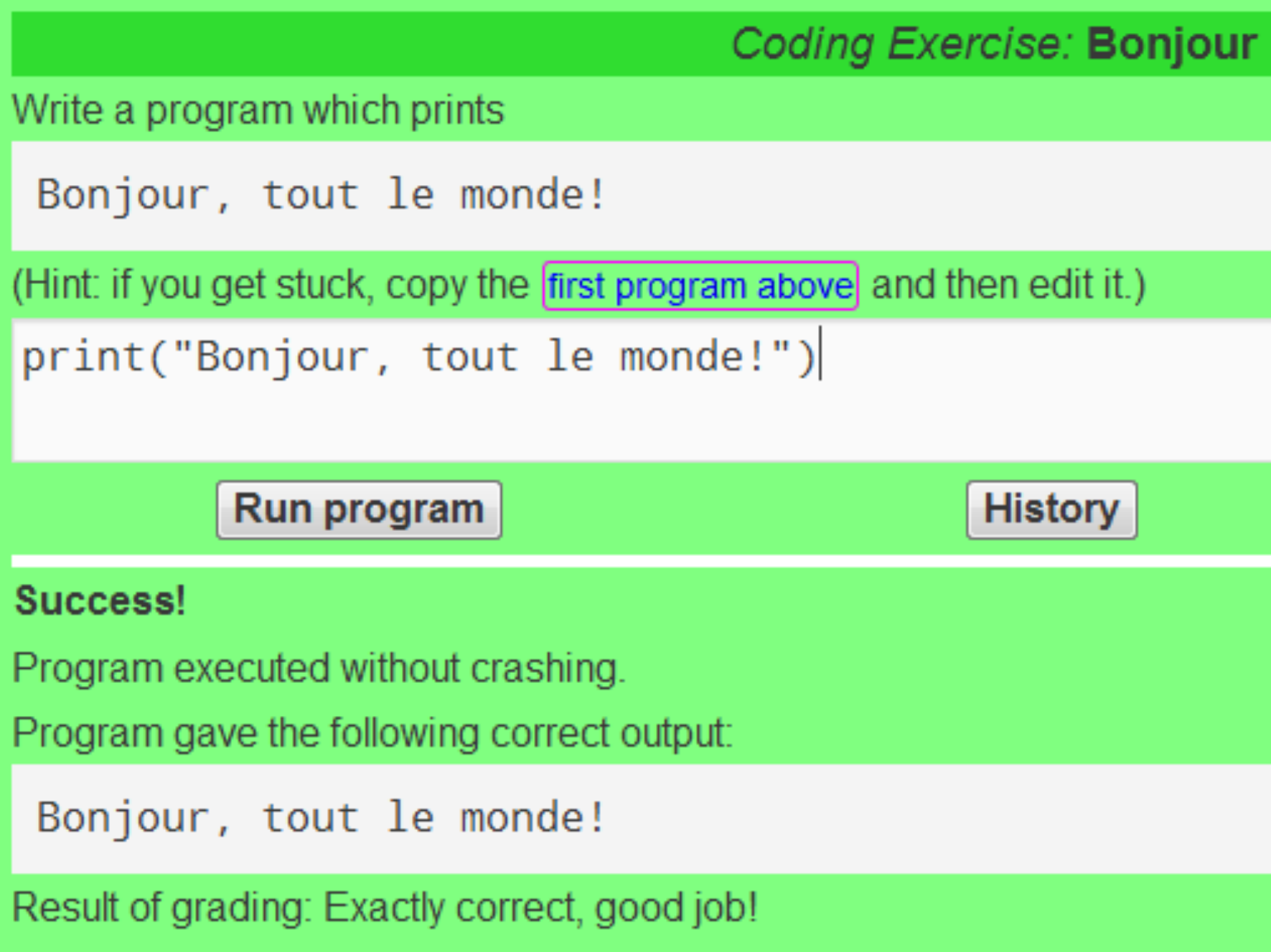}
\caption{Top: the first exercise in CS Circles. Bottom: when the user submits a correct submission, the grader reply is shown (without a page reload). The box color has changed from blue to green. A checkmark (not shown) also appears in the top right corner.}\label{fig:hello}
\end{figure}

Many of the sites we have mentioned use ``gamification" features like points and badges (see also \url{www.checkio.org}). CS Circles has little of this since we believe that it can be intimidating for beginners, especially when not all users have access to the same content and features. The full extent of gamification in CS Circles is that when an exercise is completed, its box turns green and a checkmark appears. The ``My Progress" page keeps a hyperlinked visual list of which exercises are completed, so students can easily pick up where they left off or go back over previously-skipped exercises. We receive a lot of positive user feedback; here are portions of unsolicited quotes from two students about their enjoyment of learning Python and solving the exercises.

\begin{itemize}
\item[Jun] 11 2012: ``\emph{I'm addicted to how the frustration of getting the function or program wrong at first transitions into excitement as I finally figure it out, and it all makes perfect sense laid out in front of me.}"
\item[Aug] 14 2012: ``\emph{Sites like Codecademy tend to corral you through a safe introduction of `type this after me' but Computer Science Circles has really helped me apply the things I'm learning in a fun, more thought-intensive way.}"
\end{itemize}
Our position is that games are addictive because they are fun and engaging, and that copying the fun and engagement is more effective than copying more superficial aspects. Likewise, because we assume that anyone completing our exercises is doing so just for the pursuit of learning and/or problem-solving, we don't take any precautions against plagiarism. This means that in an institutional setting, while CS Circles can be used to deliver content, its exercises should not be used as a way of assigning marks.

\subsection{Usage Statistics}
\begin{figure}
\centering
\begin{tikzpicture}
\begin{semilogyaxis}[
log basis y=2,
ytick={64,128,256,512,1024,2048,4096},
yticklabels={64,128,256,512,1024,2048,4096},
ymin=64,
only marks,
grid=major,
xtick={27,53,79},
xticklabels={,,},
title={All 105 problems (sorted by most to least solved)},xlabel={},ylabel=Completed by this many people,
enlargelimits=false]
\addplot file {completersrevised.txt};
\end{semilogyaxis}
\end{tikzpicture}

\vspace{2.5mm}

\begin{tikzpicture}
\begin{semilogyaxis}[
ytick={1,4,16,64,256,1024,4096,8192},
yticklabels={1,4,16,64,256,1024,4096,8192},
x tick label style={/pgf/number format/1000 sep=},
xtick={1000,2000,3000,4000,5000,6000,7000},
only marks,
grid=major,
title={Registered users (sorted by most to least submissions)},
xlabel={},
ylabel=Submitted code this many times,
enlargelimits=false]
\addplot file {submissionsrevised-arxiv.txt}; 
\end{semilogyaxis}
\end{tikzpicture}

\vspace{4mm}

\begin{tikzpicture}
\begin{semilogyaxis}[
ymin=0.02,
ytick={0.0416,0.2083,1,4,14,56},
yticklabels={1 hr,5 hrs,1 day,4 days,2 wks,8 wks},
grid=major,
xtick={27,53,79},
xticklabels={,,},
log basis y=2,
only marks,
xticklabels={,,},
title={All 105 problems (in order of appearance)},
xlabel={},
ylabel=Time until solved (octiles),
enlargelimits=false]
\addplot[color=black, mark=*, mark size=1pt] table[x=problem,y=1] {octiles.txt}; 
\addplot[color=red, mark=*, mark size=1.5pt] table[x=problem,y=2] {octiles.txt}; 
\addplot[color=black, mark=*, mark size=1pt] table[x=problem,y=3] {octiles.txt}; 
\addplot[color=blue, mark=*] table[x=problem,y=4] {octiles.txt}; 
\addplot[color=black, mark=*, mark size=1pt] table[x=problem,y=5] {octiles.txt}; 
\addplot[color=red, mark=*, mark size=1.5pt] table[x=problem,y=6] {octiles.txt}; 
\addplot[color=black, mark=*, mark size=1pt] table[x=problem,y=7] {octiles.txt}; 
\end{semilogyaxis}
\end{tikzpicture}
\caption{Top: The number of people who have completed each problem. Middle: The number of times each user of the site has submitted code. Bottom: For each problem, conditional on solving it, octiles of time elapsed between registration and completing it.}
\label{fig:data}
\end{figure}

In \prettyref{fig:data} we show three different charts describing usage data for CS Circles. The top chart shows the number of users that have completed each of the 105 problems currently on the site, sorted from most completed to least completed. The ``outlier" with the least number of completions is the final exercise on the site, which asks the user to look up information from Wikipedia and efficiently compute a list of primes by implementing the Sieve of Eratosthenes.

The middle chart in \prettyref{fig:data} shows the total number of code fragments submitted by each user. Note that some users have submitted several thousand code fragments.

The bottom chart in \prettyref{fig:data} shows the typical time elapsed (in octiles) before each problem is completed. The completion time for each problem is approximately normally-distributed over users. Interestingly, for \emph{every} exercise, at least half of all students who ever complete it do so within two weeks.

We have no formal biometric data, but in unsolicited feedback and the help system, a few users have self-identified as a PhD student, a lawyer, a parent and child, a miner working in a remote location, and individuals over age 60. CS Circles can help a school reach out to non-traditional student populations, including ``math circles"-style settings~\cite{kaplan2007} where one might assign online exercises between weekly meetings.

\section{Features}\label{sec:features}

\subsection{Student Tools}\label{sec:studenttools}
While a typical IDE has dozens of buttons, our coding windows start with just 3: submit, help, and history. The remaining features are explicitly introduced and explained one at a time in later sections, so that the student learns the purpose of all the available tools. This technique helps avoid novices from getting overwhelmed.

The ``Help" button is for the student to use when they are stuck and want targeted assistance. Clicking on this button opens up a new text field where the student is directed to type a few sentences about what they've tried and where they are stuck. This is sent along with the current contents of the code editor for that problem.

For each user, every coding exercise automatically stores a history of all prior submissions. The history can be accessed on demand, both by the student and their guru. When a user re-visits a page, the most recently submitted version is automatically loaded in the text editor.

In lesson 4, we introduce the two debugging tools for the course. One is simply a ``console" that allows the student to run arbitrary code without any grader interference. Every example has a button which allows it to be copied to the console, in a new browser tab, with a single click. The other debugging tool is an open-source Python visualizer~\cite{Guo11} that can take an arbitrary program, store the steps of its execution, and replay those steps both forwards and backwards in time along with a visual representation of all the variables and the call stack. This replicates most of the functionality of a client-side debugger but the backwards execution is even more useful than what most IDEs currently offer. Students run the visualizer on 650 pieces of code per day on average.

The ability to specify a test input is introduced in lesson 5, when the command to read from standard input is introduced. In later lessons, when users are tested based on their function definitions, this input box is replaced with a text area for test arguments.

In lesson 7, just after introducing the first block-structured elements of Python, we switch the user to a ``rich" editor, specifically the CodeMirror editor~\cite{Haverbeke1x}. It highlights syntax, numbers lines, matches parentheses, and performs smart indentation. We have customized it to resize automatically up to a fixed limit, but to also be manually sizeable when the user desires.

\subsection{Lesson Designer Tools}
CS Circles is exercise-centric: since we believe you learn best by active participation, exercises are embedded throughout the lessons. We use traditional exercise formats like short answer and multiple choice as well as many varieties of coding exercises.

The Python auto-grader is similar to existing auto-graders used by olympiads and public problem websites. Aiming for beginners necessitates that the grader offer more grading styles than just the typical stdin/stdout approach. While we teach the {\tt print} function in lesson 0, input reading doesn't occur until lesson 5. For this reason the early exercises specify input through variables that the grader pre-populates with values. In lesson 1, the \emph{swap} exercise --- write code to put the value of {\tt x} in {\tt y} and vice-versa --- has the grader pre-define variables, and test their values after the user code executes. Later, once we have described how to define functions with {\tt def}, inspecting the values returned by function calls becomes the main testing approach. We believe that each of these three methods (variable-based, stdin/stdout, function-based) has its place in a beginner course.

Our grader uses a number of other techniques in order to maximize the variety and fun of the exercises:
\begin{itemize}
\item randomizing test cases, with a model solution to generate answers, to prevent hard-coded solutions;
\item pre-defining non-working initial code --- additionally, sometimes we limit the (Levenshtein) edit distance to a valid solution;
\item pre-defining a function for the student;
\item forbidding the student from using some built-in functions;
\item requiring that a specific error be produced;
\item and plugging in custom graders for problems where more than one answer is correct.
\end{itemize}

{\bf Code Scramble.} We present a small fraction of our coding exercises as \emph{code scramble} exercises: a solution is already given in the coding window, but with its lines appearing in the wrong order; users are only permitted to drag-and-drop entire lines of code until they find  a correct solution. A nice side effect of writing a problem in this way is that we don't rely on the student to have memorized any past syntax, and they can focus purely on the logical aspect of the problem. See also the studies~\cite{LWRL08,FT09} on ``syntax-free programming."
The problem type we call ``code scramble" has been previously called \emph{Parsons problems}~\cite{PH06}; see Denny et al.~\cite{DLRS08} for a thoughtful quantitative study of how complex these problems should be for optimal effect. One interesting feature suggested previously that we have not implemented is to have extra lines that must be left out of the solution. Generally, our code scrambles are of medium difficulty, and syntactic considerations alone will partially, but not completely, solve them.
See \cite{HLP11} for information on \emph{C-doku}, near-complete programs with specific short answer blanks that the student must fill in, with additional features such as code coverage requirements.

{\bf Hints} are essential to writing an exercise that can meet the needs of the widest variety of students. As a rule, our problems are always solvable using only the tools that have been previously introduced to the student, even without hints. However, we also want to avoid students getting stuck on a particular problem just because they have overlooked an old idea or they are unable to come up with an innovation on their own. For this reason we offer clickable hints. The most basic version is one whose content appears like a pop-up window when clicked, which can be moved around the page to a convenient place and closed when no longer needed. Occasionally we find that a very conceptually challenging exercise (such as the swap exercise) requires many hints, and in this case we present them in an ``accordion" style, where at most one of the hints can be slid open on the page in the midst of the existing content, which avoids overloading the user interface with too many boxes.

Two other features are the result of user feedback: a ``cheatsheet" in html and pdf formats, which contains a quick review of all the functions and syntax used by our curriculum; and ``previous/next lesson" buttons on the bottom of each page, precluding the need to scroll all the way up to the main navigation menu when a lesson has been completed.

\subsection{Teacher Tools}
A student can ask any teacher, mentor or friend who has signed up for CS Circles to be their ``guru." E.g., in any class using CS Circles the teacher would normally get the students to enter him/her as their guru. Currently, there are 25 gurus that have five or more students registered under them; in total these gurus have almost 400 registered students.

A student with a guru can send their ``Help" messages to their guru instead of to the CS Circles staff. When a teacher clicks on the reply link sent in the e-mail that they receive, they are brought to the ``mail" webpage. The mail page offers the teacher several tools we found invaluable in providing fast personalized responses:
\begin{itemize}
\item the complete submission history for that student on that problem, to see what they've tried that worked or did not, and how close they got;
\item a view of the ``My Progress" page for that student, to see where they are overall;
\item and a list of the teacher's other messages about the same problem, which allows them to cut-and-paste relevant information when several different students have similar questions.
\end{itemize}
We have found that this system is an extremely efficient way to provide targeted, personalized feedback to hundreds of students without much work, filling in the inherent gaps in what an auto-judge can provide.

There is another surprising benefit of this ``Help" button: as lesson authors, we are able to get a tremendous amount of quick feedback when any newly-introduced content is too difficult or badly worded. Many of the hints in the current curriculum are the result of boiling down commonly-asked questions received in this way.

\usanon{}{\eject}
\section{Implementation}\label{sec:implementation}
CS Circles is built on top of the WordPress content management system (CMS). It is estimated\footnote{\url{http://trends.builtwith.com/cms}} to be the most popular CMS, hosting a significant portion of all web traffic. We chose it because it has an easy web-based interface for management and content editing (including instant previews), it has a built-in secure login system we could extend for our own purposes, it is built on a well-maintained and easy-to-modify code base, and it has a plugin system which lets us easily add third-party features. We use plugins for site search, analytics, contact forms, translation, \LaTeX, jQuery UI, and translator/admin role management.

As mentioned earlier, CodeMirror~\cite{Haverbeke1x} and the Python visualizer~\cite{Guo11} are two prominent non-WordPress tools that we integrated into CS Circles. Flexigrid\footnote{\url{http://flexigrid.info/}} is another external library that we use, which allows for smooth and secure user access to database slices --- for example this lets you browse through your past code submissions without pulling them all from our server at once.

Lesson pages are stored, like in most CMSes, as html-formatted text in a database. The auto-grader and exercises are specified using WordPress ``shortcodes." For example, the code at the top of \prettyref{fig:heads} defines an exercise where students must correctly define variables counting the heads, shoulders, knees, and toes at a party, given a pre-defined count of the number of {\tt people}.
The specification, when embedded in the lesson text and rendered, produces the exercise. At the bottom of \prettyref{fig:heads} we show an example of the grader's reply to a partially correct solution. This quick text-based specification system has been instrumental in quick prototyping and lesson editing.

\begin{figure}
\centering
\begin{minipage}{10cm}
\begin{verbatim}
[pyBox repeats=3 precode="people = _rint(10, 100)"
autotests="heads\nshoulders\nknees\ntoes"
solver="heads, shoulders, knees, toes =
people, 2*people, 2*people, 10*people"]
\end{verbatim}
\end{minipage}
\vspace{0em}$$\Downarrow$$
\includegraphics[width=8cm]{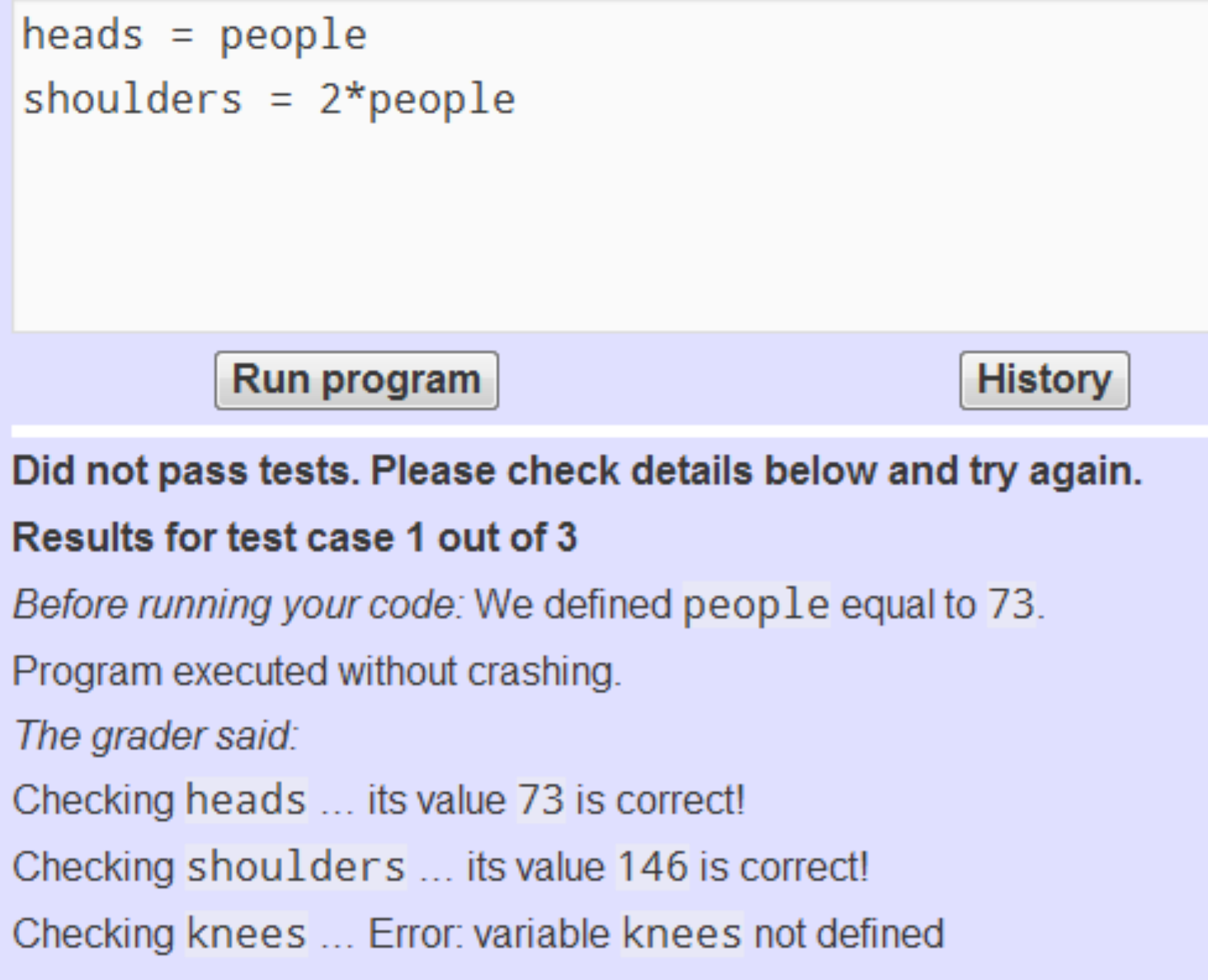} 
\caption{Top: a WordPress-style shortcode that defines an exercise in CS Circles. Bottom: the result of a submission sent to this exercise.}\label{fig:heads}
\end{figure}

The auto-grader and surrounding code is written in a mix of several languages. There is html/js/css to handle browser interaction, PHP to interface with WordPress and generate dynamic html, Python to do introspective grading, and C++ to call the core ``safeexec" sandbox routines. The latter was developed starting from the {\tt safeexec} module of the Mooshak contest system~\cite{LS03} --- our fork is available at \url{http://github.com/cemc/safeexec}. For a survey of the techniques involved in executing arbitrary user code securely, see the excellent survey of Fori{\v s}ek \cite{Forisek06} or the recent implementation of Mare{\v s} and Blackham~\cite{MB12}.

The grader has a run-time limit of 1 second by default. With the sole exception of the Sieve of Eratosthenes exercise, we don't intend that any of the coding problems need to be solved in an optimized way. So far the design has been successful: we've not seen any correct student code that times out on any other problem.


\section{Future Work}\label{sec:future}
Our experiences with WordPress and its great plugin community drive us to ask, would it be useful to distribute the technologies behind CS Circles as a set of plugins? It would be excellent to let individual educators develop and host their own interactive programming lessons.

We would like to extend the technology behind the site to other languages such as JavaScript, R, and Sage. Additionally, it would be beneficial to integrate interactive sandboxes (and exercises) into programming languages' online documentation.

Interactive input and output is not currently possible in CS Circles, but it would be an important feature to add in a future version. The Skulpt environment used by \cite{MR11,MR12} would be one way to accomplish this.

While programming generally requires enough English competency to make sense of the syntax and the errors, we have translated everything else (including the UI and grader feedback) to French. Translations into a few other languages are under way, and reaching more students and teachers in this way is an exciting prospect.

Can a useful programming course be taught on a touch-screen/tablet/iPad? Our intuition is that a comfortable keyboard is an important ingredient in making any significant amount of coding enjoyable. However, ``code scrambles" might work even better on a touchscreen.

\section{Acknowledgments}
\usanon{
We thank our colleagues Graeme Kemkes, Sandy Graham, and Andy Kong for helping with lesson/backend development, answering student questions, and server administration, and Brice Canvel for leading the French translation. We thank the authors of the plug-ins and libraries we use, especially Philip Guo and Peter Wentworth for building the visualizer, and the authors of the Polylang plugin. We are inspired by StarCraft's excellent integrated tutorials and choosable level ordering, and by \emph{Good Eats} for explanations that are at once clear, amusing, scientific, and polished.}{[[redacted]]} 

\eject\bibliographystyle{abbrv}
\bibliography{cscircles}

\end{document}